\documentclass[runningheads]{llncs}
\usepackage{subfig}
\usepackage{adjustbox}
\usepackage{makeidx}
\usepackage{float}
\usepackage{multirow}
\usepackage{subfig}
\usepackage{graphicx}
\usepackage{booktabs}
\usepackage{xcolor}
\usepackage{amsmath}
\usepackage{amsfonts}
\usepackage{caption} 
\begin{document}
\title{Structurally aware bidirectional unpaired image to image translation between CT and MR}
%
%


\author{Vismay Agrawal \and
Avinash Kori \and
Vikas Kumar Anand\and 
Ganapathy Krishnamurthi}

\authorrunning{V. Agrawal et al.}
\institute{Department of Engineering Design, Indian Institute of Technology Madras, India\\
\email{gankrish@iitm.ac.in}}

\maketitle      

\begin{abstract}
Magnetic Resonance (MR) Imaging and Computed Tomography (CT) are the primary diagnostic imaging modalities quite frequently used for surgical planning and analysis. A general problem with medical imaging is that the acquisition process is quite expensive and time-consuming. Deep learning techniques like generative adversarial networks (GANs) can help us to leverage the possibility of an image to image translation between multiple imaging modalities, which in turn helps in saving time and cost. These techniques will help to conduct surgical planning under CT with the feedback of MRI information. While previous studies have shown paired and unpaired image synthesis from MR to CT, image synthesis from CT to MR still remains a challenge, since it involves the addition of extra tissue information. In this manuscript, we have implemented two different variations of Generative Adversarial Networks exploiting the cycling consistency and structural similarity between both CT and MR image modalities on a pelvis dataset, thus facilitating a bidirectional exchange of content and style between these image modalities. The proposed GANs translate the input medical images by different mechanisms, and hence generated images not only appears realistic but also performs well across various comparison metrics, and these images have also been cross verified with a radiologist. The radiologist verification has shown that slight variations in generated MR and CT images may not be exactly the same as their true counterpart but it can be used for medical purposes.

\end{abstract}
\section{Introduction}
\par Computer Tomography (CT) and Magnetic Resonance (MR) Imaging are the two most frequently used imaging modalities for medical diagnosis and surgical planning. While CT provides contrast information for bone studies, MR imaging provides information for soft tissue contrast. In this manuscript we propose a Generative Adversarial Networks (GANs) \cite{goodfellow2014generative} based technique to convert CT images to realistic looking MR images. As the physics of acquisition in the case of MR and CT differ significantly, developing techniques for direct transformation of images from MR space to CT space is not a trivial problem. But recent advancements in data-driven modeling, specifically GANs have provided realistic looking results on image to image translation by separating content and style from an image. We make use of similar techniques to separate style and content from MR and CT images and interchange them. To overcome the challenges in obtaining paired data, we conducted our experiments along the lines of cycleGAN \cite{zhu2017unpaired} based approach, where the models were trained on unpaired data to maintain the cyclic consistency between different imaging modalities. The use of unpaired data drastically increases the amount of input-output pairs used in translational networks. It also assures that the network is not overfitting itself on the given data but is learning the visual properties of different image modalities. We have also played with loss functions and have shown that using structural similarity in loss function can generate realistically looking images which can be used in looking at bone structure and fractures in a better way. These findings were also cross verified by a radiologist.

Previously, Conditional GANs \cite{isola2017image} have been used with U-net \cite{Ronneberger2015UNetCN} based architecture as a generator and convolutional PatchGAN classifier as a discriminator to learn the mapping from input images to output images. In some instances, CNNs and GANs have also been implemented in MR to CT translation \cite{kaiser2019mri} \cite{nie2017medical}. cycleGANs have been used to perform a bidirectional unpaired image to image translation between Cardiac CT and MR data \cite{chartsias2017adversarial} and it has been shown that the generated data can be used to increase the accuracy of the network in segmentation tasks. Wolterink et al. \cite{wolterink2017deep} has shown that cycleGAN trained on unpaired data was able to outperform the model trained on paired data for MR to CT translation. Jin et al. \cite{jin2019deep} study was focused on CT to MR translation which consisted of dual cycle-consistency loss using paired and unpaired training data, but due to the use of paired data in the training process, the method can't be used in the situation with a lack of paired data. With all these, we can see that although many research work has been done on brain and cardiac medical scans, there aren't many successful works with pelvis scans. Also, we have observed that most of the research was focused on obtaining CT images from their MR counterpart, with little or no details about bi-directional translations. 

\section{Materials and Methods}

\subsection {Data}
Our data consisted of human pelvis unpaired (taken from multiple patients) MR and CT 3D volumes. T2 weighted images were obtained from local hospital (Apollo Speciality Hospital, Chennai, TamilNadu, India) with appropriate ethical clearance which were acquired using a 1.5 Tesla system (Achieva, Philips Medical System), while CT volumes we extracted from Liver Tumour Segmentation (LiTS) challenge database \cite{LITS}, under the guidance of expert radiologists. In total, we have about 55 MR and CT volumes, average with about 30 slices in MR and about 80 slices in CT volumes. We extracted axial slices from these MR and CT volumes to train the network. The data split used in training and testing is shown in table \ref{Data}.

\begin{table}[h]
    \centering
    \setlength{\tabcolsep}{0.5em}
    \caption{Dataset Details}
    \label{Data}
    \begin{tabular}{@{}lcccc@{}}
        \toprule
                  & Training Set & Test Set & Volume Dimension &  \\ \midrule
        CT Volume & 50           & 5        & (512, 512, 80)   &  \\
        MR Volume & 50           & 5        & (768, 768, 30)   &  \\ \bottomrule
    \end{tabular}
    
\end{table}

\subsection{Pre-processing of data}
As the number of slices was different in MR and CT, to maintain an equal number of slices we resampled all volumes to have 80 slices on an average. Each image slice was then normalized using the min-max normalization method. The obtained image was resized to the dimension (286, 286) using bicubic interpolation and subsequently, it was cropped to a dimension (256, 256) from a random location. Data augmentation was performed by flipping and rotating the input image at random.

\subsection{Transformer Network}
\begin{figure}[h]
    \centering
    \includegraphics[width = 1.0 \textwidth]{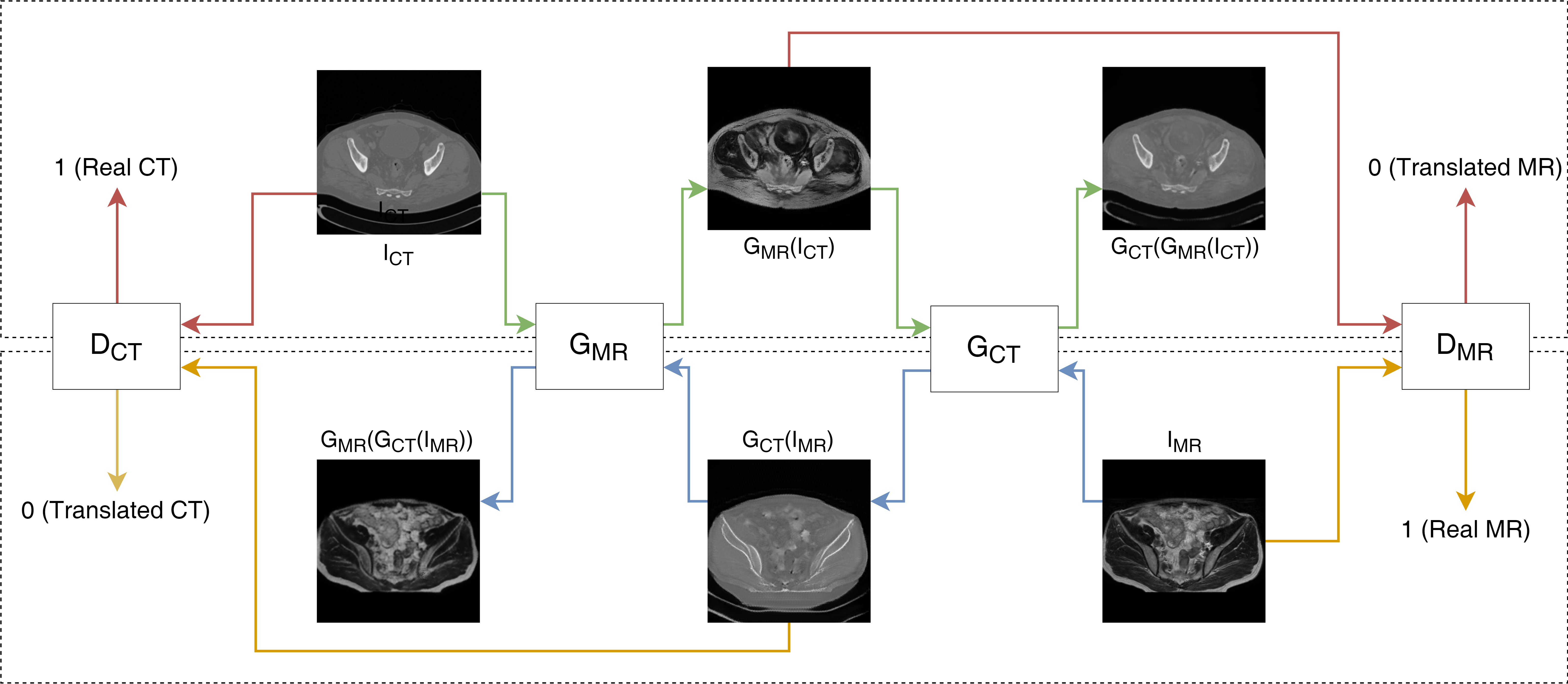}
    \caption{Schematics of our method used inter sequence MRI images}
    \label{pipeline}
\end{figure}

Our pipeline consisted of two identical Generator networks for CT to MR and MR to CT image to image translation, i.e. $G_{CT}: I_{MR} \rightarrow{} I_{CT}$ and $G_{MR}: I_{CT} \rightarrow{} I_{MR} $. During training, one of the networks performs the required translation and the other one tries to undo the same. We have used ResNet \cite{he2016deep} as a generator network with 9 resblocks. For each generator, we have a corresponding discriminator network,  $D_{CT}$ and $D_{MR}$, which aims to discriminate between translated and original, CT and MR images respectively. The pipeline of our network is given in Fig. \ref{pipeline}

\subsection{Loss function}
In the two proposed networks, the first one (cycleGAN) uses loss function as proposed in \cite{Yang2018UnpairedCycleGAN} and in the second network (cycleGAN-SSIM) we have included the  structural similarity loss function \cite{wang2004image} on top of all the other losses. The loss functions can be divided into multiple parts as stated below. 

\subsubsection{GAN Loss}
\begin{align*}
    \mathcal{L}_{GAN} (D_{MR}, G_{MR}, I_{CT})= \ & \mathbb{E}_{I_{CT} \sim Data_{CT}}[(D_{MR}(G_{MR}(I_{CT})) - 1)^2] \\ 
    & + \mathbb{E}_{I_{MR} \sim Data_{MR}}[(D_{CT}(G_{CT}(I_{MR})) - 1)^2]
\end{align*}
The GAN loss aims to minimize the possibility of the translated image being recognized as a fake image by the discriminator.

\subsubsection{Cycle Loss}
\begin{align*}
    \mathcal{L}_{cycle} = \ & \mathbb{E}_{I_{MR} \sim Data_{MR}}[||G_{MR}(G_{CT}(I_{MR})) - I_{MR})||_1] \\ 
    & + \mathbb{E}_{I_{CT} \sim Data_{CT}}[||G_{CT}(G_{MR}(I_{CT})) - I_{CT})||_1]
\end{align*}
The cycle loss function ensures the cyclic consistency between the input and generated images.

\subsubsection{Identity Loss}
\begin{align*}
    \mathcal{L}_{identity} = \ & \mathbb{E}_{I_{CT} \sim Data_{CT}}[||(G_{CT}(I_{CT})) - I_{CT})||_1] \\
    & + \mathbb{E}_{I_{MR} \sim Data_{MR}}[||(G_{MR}(I_{MR})) - I_{MR})||_1]
\end{align*}
The root of the identity loss lies in the fact that the generator should be aware of the input image modality, and the generator shouldn't translate the input image if it corresponds to same imaging modality as the output of the generator.

\subsubsection{Structural Similarity (SSIM) Loss}
\begin{align*}
    \mu_{MR/CT} &= \mathbb{E}_{I_{MR/CT} \sim Data_{MR/CT}}[G_{CT/MR}(I_{MR/CT})] \\
    \sigma_{MR/CT} &= \mathbb{E}_{I_{MR/CT} \sim Data_{MR/CT}}[(G_{CT/MR}(I_{MR/CT}))^2]\\
    \sigma_{MR, CT} &=  \mathbb{E}_{I_{CT} \sim Data_{CT}, I_{MR} \sim Data_{MR}}[G_{MR}(I_{CT})G_{CT}(I_{MR})]\\
    \mathcal{L}_{SSIM} &= 1 - \ \frac{(\mu_{CT}\mu_{MR} + c_1)(2\sigma_{MR,CT} + c_2)}{(\mu_{MR}^2 + \mu_{CT}^2 + c_1)(\sigma_{MR}^2 + \sigma_{CT}^2 + c_2)}
\end{align*}

Here, we chose $c_1$ and $c_2$ to be 0.0001 and 0.009 respectively. SSIM helps in maintaining the structural integrity between inter-domain images.

\subsubsection{Generator Net Loss}
\begin{align*}
\mathcal{L}_{G_{net}} = \ & \mathcal{L}_{GAN} + \lambda_{cyc} * \mathcal{L}_{cycle} + \lambda_{id} * \mathcal{L}_{identity} + \lambda_{ssim} * \mathcal{L}_{ssim}    
\end{align*}

Net generator loss is defined as a superposition of all the aforementioned losses namely, GAN loss, cycle loss, identity and SSIM loss.

\subsubsection{Discriminator Loss}

\begin{align*}
    \mathcal{L}_{dis_{MR/CT}} = &\ \mathbb{E}_{I_{CT/MR} \sim Data_{CT/MR}}[\ (D_{MR/CT}(I_{CT/MR}) - 1)^2\ ] \\
     &+ \mathbb{E}_{I_{CT/MR} \sim Data_{CT/MR}}[\ (D_{MR/CT}(G_{MR/CT}(I_{CT/MR}))^2\ ]
\end{align*}

As our aim is to predict label 1 for a real imaging modality and label 0 for a translated one, the loss function is the distance between the prediction and the real label.

\subsection{Comparison Metrics}
Quantitative analysis of the generated images was performed using FID index (Fréchet Inception Distance) \cite{gan_metric}, MI (mutual information) \cite{gan_metric} and SSIM index \cite{gan_metric}. 

\subsubsection{FID}
In case of FID, pretrained densenet121 ($FID(.)$) is considered as base network, and feature embedding for real MRI/CT and corresponding fake MRI/CT is obtained, the distance between these two embeddings are observed. If the distribution of generator matched with the distribution of real image the distance between feature embeddings will reduce and FID score will increase. 

\begin{align*}
    {FID}_{CT} = &\ \mathbb{E}_{I_{CT} \sim Data_{CT}, I_{MR} \sim Data_{MR}}<FID(G_{CT}(I_{MR})), FID(I_{CT})>
\end{align*}
where $ <.> $ indicates inner product between two vectors. FID for MR can be calculated similarly.
%
\subsubsection{SSIM}
To quantitatively measure the structural integrity of the generated images we made use structural similarity as a metric. SSIM between generated CT/MR and input MR/CT provides us with an information how structurally close the given pair of generated MR and CT are.
\begin{align*}
    {SSIM} &= \frac{(\mu_{CT}\mu_{MR} + c_1)(2\sigma_{MR,CT} + c_2)}{(\mu_{MR}^2 + \mu_{CT}^2 + c_1)(\sigma_{MR}^2 + \sigma_{CT}^2 + c_2)}
\end{align*}

\subsubsection{MI}
To measure the textural similarity between generated MR/CT with actual MR/CT we made use of mutual information as a metric.
\begin{equation*}
\label{MI}
MI(G, R) = \mathbf{E}(P_{GR}(G,R)) \times log(\frac{P_{GR}(G,R)}{P_{G}(G)P_{R}(R)})
\end{equation*}
where $P_{GR}(G, R)$ denotes joint distribution between generated and real distribution, $\mathbf{E}$ denotes expectation value and $P_G(G), P_R(R)$ denotes marginals.

\subsubsection{Pixel wise accuracy (pixacc)}
To gauge the performance of transformation network in generating recovered MR/ CT using real MR/ CT images, cosine similarity between them was used as a metric, as stated below.
\begin{equation*}
\label{pixacc}
pixacc_{MR/CT} = \mathbb{E}_{I_{CT} \sim Data_{CT}, I_{MR} \sim Data_{MR}}\frac{I_{MR/CT} \cdot G_{MR/CT}(I_{CT/MR})} {||I_{MR/CT}||\cdot ||G_{MR/CT}(I_{CT/MR})||}
\end{equation*}

\section{Results and Discussion}
\begin{figure}[h]
\centering
\subfloat[Total Generator loss]{\includegraphics[width=0.5\textwidth]{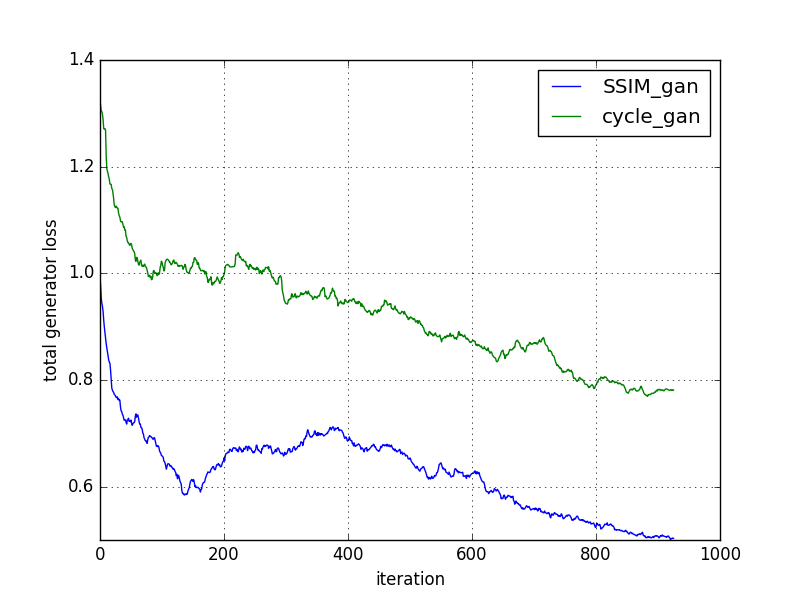}}\hfill
\subfloat[Total Discriminator loss]{\includegraphics[width=0.5\textwidth]{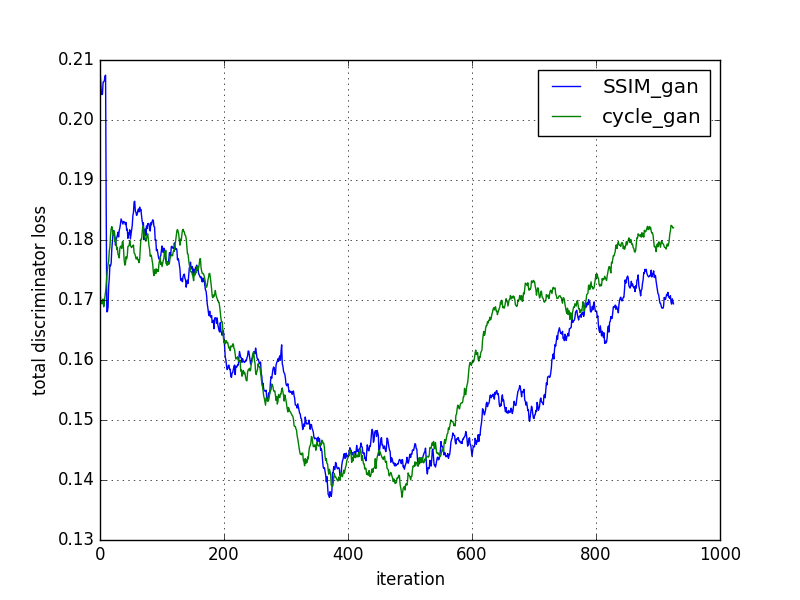}}
\caption{comparison of models based on loss convergence}
\label{loss}
\end{figure}

The performance metrics were being calculated for 200 slices and the average results are shown in Table \ref{scores}. The loss plots between cycleGAN and cycleGAN-SSIM, given in Fig \ref{loss} clearly show the lower convergence in case of using structural similarity loss for both generator and discriminator, we can conclude that the network was able to exploit the structural consistencies between different image modalities to boost its learning rate.

\begin{table}[h]
\centering
\setlength{\tabcolsep}{0.5em}
\renewcommand{\arraystretch}{1.2}
\caption{Mean comparison metrics }
\label{scores}

\begin{tabular}{l|cccc|cccc}
\hline
\multicolumn{1}{c|}{} & \multicolumn{4}{c|}{CT to MR Translation} & \multicolumn{4}{c}{MR to CT Translation} \\
Model                 & FID   & SSIM  & MI     & pixacc  & FID  & SSIM   & MI    & pixacc  \\ \hline
cycleGAN              & 0.193 & 0.408 & 0.273  & 0.986   &  0.177 & 0.562 & 0.332 & 0.994   \\
cycleGAN-SSIM         & 0.200 & 0.416 & 0.290  & 0.988   &  0.184 & 0.562 & 0.336 & 0.995   \\ \hline
\end{tabular}
\end{table}

\begin{figure}[]
\centering
\subfloat{\includegraphics[width=0.195\textwidth ]{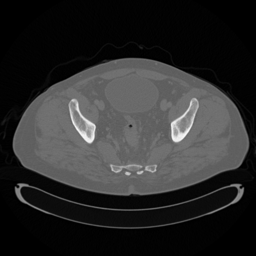}}\hfill
\subfloat{\includegraphics[width=0.195\textwidth] {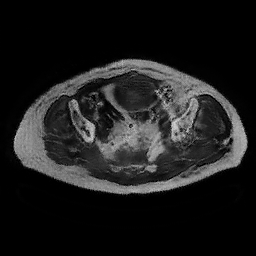}}\hfill
\subfloat{\includegraphics[width=0.195\textwidth] {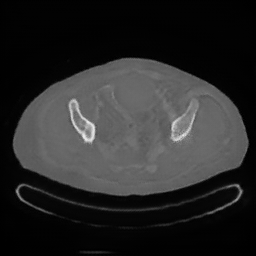}}\hfill
\subfloat{\includegraphics[width=0.195\textwidth] {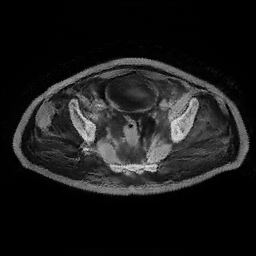}}\hfill
\subfloat{\includegraphics[width=0.195\textwidth] {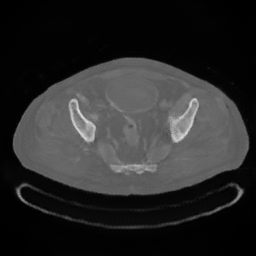}}\\
\vspace{-9pt}

\subfloat{\includegraphics[width=0.195\textwidth ]{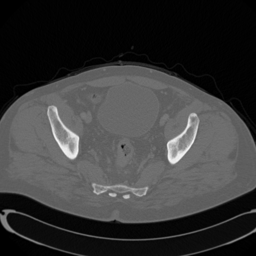}}\hfill
\subfloat{\includegraphics[width=0.195\textwidth] {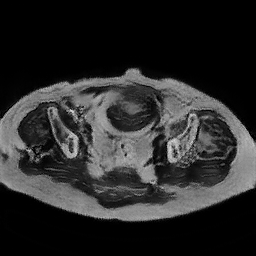}}\hfill
\subfloat{\includegraphics[width=0.195\textwidth] {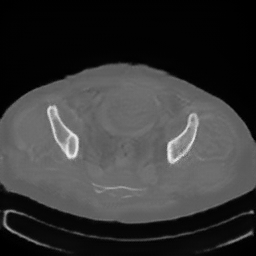}}\hfill
\subfloat{\includegraphics[width=0.195\textwidth] {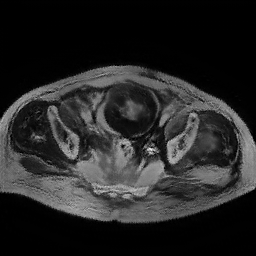}}\hfill
\subfloat{\includegraphics[width=0.195\textwidth] {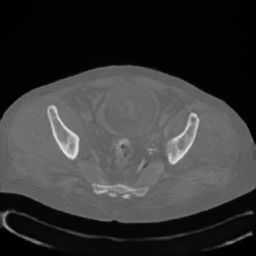}}\\
\vspace{-9pt}


%
\subfloat{\includegraphics[width=0.195\textwidth ]{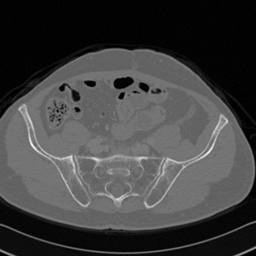}}\hfill
\subfloat{\includegraphics[width=0.195\textwidth] {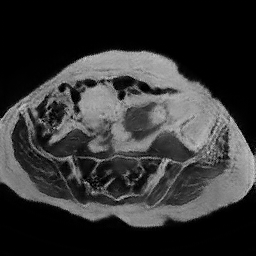}}\hfill
\subfloat{\includegraphics[width=0.195\textwidth] {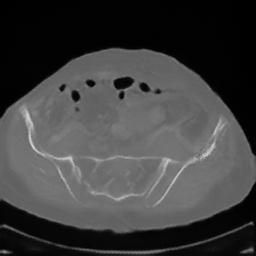}}\hfill
\subfloat{\includegraphics[width=0.195\textwidth]{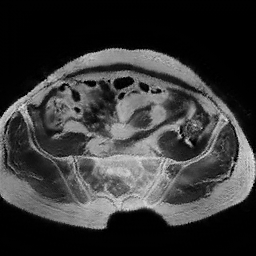}}\hfill
\subfloat{\includegraphics[width=0.195\textwidth] {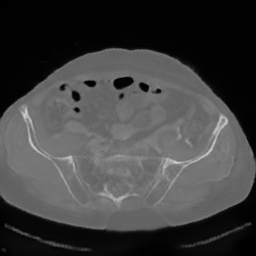}}\\
\caption{CT to MR conversion. In the above image column (i) corresponds to real CT, (ii) generated MR using cycleGAN (iii) recovery CT using cycleGAN (iv) generated MR with cycleGAN-SSMI (v) recovery CT with cycleGAN-SSMI}
\label{CT2MRexamples}
\end{figure}

\begin{figure}[]
\centering
\subfloat{\includegraphics[width=0.195\textwidth ]{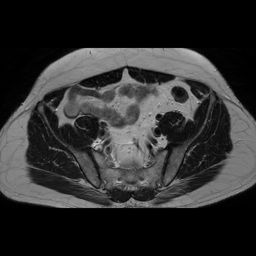}}\hfill
\subfloat{\includegraphics[width=0.195\textwidth] {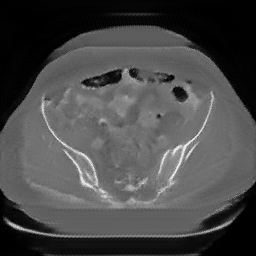}}\hfill
\subfloat{\includegraphics[width=0.195\textwidth] {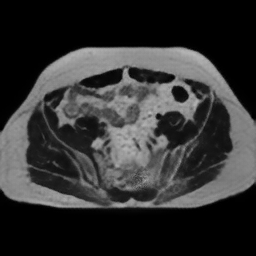}}\hfill
\subfloat{\includegraphics[width=0.195\textwidth] {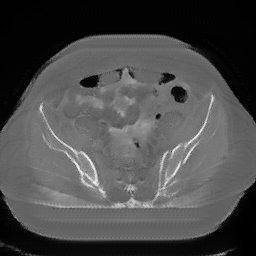}}\hfill
\subfloat{\includegraphics[width=0.195\textwidth] {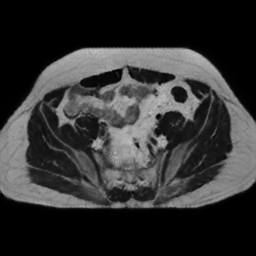}}\\
\vspace{-9pt}

\subfloat{\includegraphics[width=0.195\textwidth ]{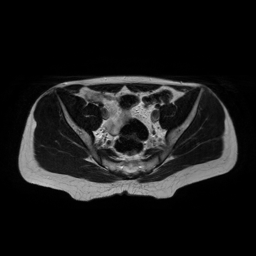}}\hfill
\subfloat{\includegraphics[width=0.195\textwidth] {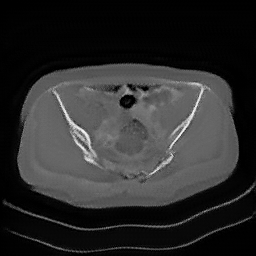}}\hfill
\subfloat{\includegraphics[width=0.195\textwidth] {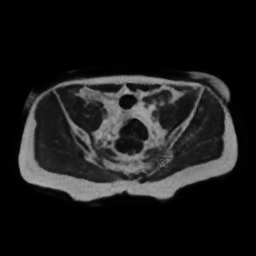}}\hfill
\subfloat{\includegraphics[width=0.195\textwidth] {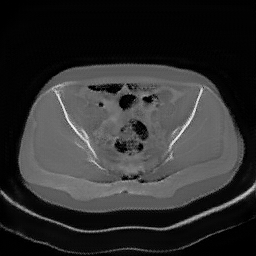}}\hfill
\subfloat{\includegraphics[width=0.195\textwidth] {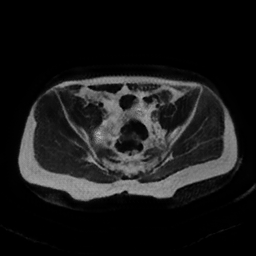}}\\
\vspace{-9pt}

%

\subfloat{\includegraphics[width=0.195\textwidth]{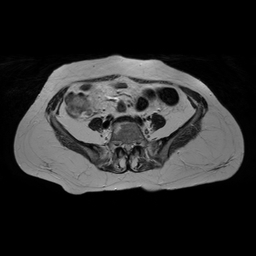}}\hfill
\subfloat{\includegraphics[width=0.195\textwidth] {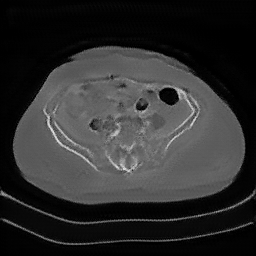}}\hfill
\subfloat{\includegraphics[width=0.195\textwidth] {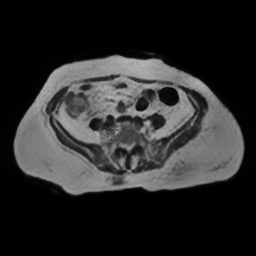}}\hfill
\subfloat{\includegraphics[width=0.195\textwidth]{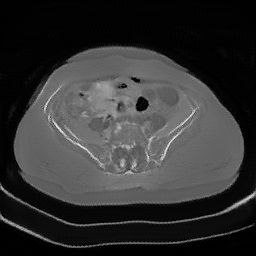}}\hfill
\subfloat{\includegraphics[width=0.195\textwidth] {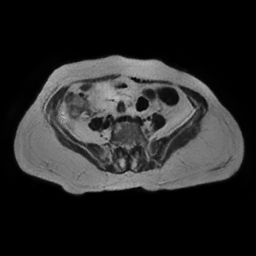}}\\
\caption{MR to CT conversion. In the above image column (i) corresponds to real MR, (ii) generated CT using cycleGAN (iii) recovery MR using cycleGAN (iv) generated CT with cycleGAN-SSMI (v) recovery MR with cycleGAN-SSMI}
\label{MR2CTexampls}
\end{figure}

We can see that although all the results were higher in the case of the network trained with SSIM loss function, there aren't any significant differences in their scores. The use of structural similarity loss function was preserving information about the edges of the generated images, especially in the case of CT to MR translation (Fig \ref{CT2MRexamples}), wherein visual inspection shows enhanced bone details, and relatively higher structural similarity and mutual information scores for cycleGAN-SSIM model. The MR to CT translation (Fig \ref{MR2CTexampls}) had similar SSIM scores for both the models, this denotes that cycleGAN was learning to structural information in CT images. In such a case addition of the SSIM loss function was just acting as a catalyst to increase the learning rate of the model. High pixacc scores reflect the fact that both the models generate realistically looking images and such synthesised images are suitable for use in medical diagnosis. MR to CT translation had higher pixacc scores than CT to MR translation because MR has more tissue contrasts and such information is missing in CT, thus it isn't possible for the network to completely recover the lost information in latter transformation.

According to the reviews by the radiologist, in comparison to cycleGAN-SSIM model, the images generated by cycleGAN were more similar to real and CT MR images. Radiologist pointed out that MR images generated by cycleGAN model would be better for routine diagnosis, whereas images generated by cycleGAN-SSIM model would be better for special cases of diagnosis. In cycleGAN-SSIM model, due to preservation in textural details of input CT image content, its MR counterpart contained contrast feature of MR image with blended with structural information of CT images. Such images can be used in parallel with CT images for diagnosis of bone fractures and also in imaging the lung to identify tiny nodules and calcification.


\section{Conclusion and Future work}
In this work,  we have shown that the cycleGAN can be used in medical image translation tasks. Moreover, the use of structural similarity loss with cycleGAN can boost the learning rate of GAN, images generated by such model are more structurally sound and hence gives a better diagnosis of fractures. In future, we plan to conduct similar experiments with entire 3D volumes, i.e 3D style and content separation from MRI and CT volumes. We would also like to extend this work to identify tiny nodule and calcification in the lungs, as pointed by the radiologist. 

\clearpage
\bibliographystyle{splncs04}
\bibliography{report.bib}
%
%

\end{document}